\documentclass[draftclsnofoot,onecolumn]{elsarticle}
\usepackage{amssymb}
\usepackage{amsmath}
\usepackage{setspace}

\usepackage{epsfig}
\usepackage{epstopdf}
\usepackage{color}

\journal{Signal Processing}
\begin{document}

\begin{frontmatter}
\title{Low-Complexity Variable Forgetting Factor Constrained Constant Modulus RLS Algorithm for Adaptive Beamforming\tnoteref{thanks}}

\author{Boya Qin}
\ead{qby666@zju.edu.cn}
\author{Yunlong Cai\corref{cor1}}
\ead{ylcai@zju.edu.cn}
\author{Benoit Champagne}
\ead{benoit.champagne@mcgill.ca}
\author{Rodrigo C. de Lamare}
\ead{rcdl500@ohm.york.ac.uk}
\author{Minjian Zhao}
\ead{mjzhao@zju.edu.cn}

\cortext[cor1]{Corresponding author}
\tnotetext[thanks]{This work was supported by the Fundamental Research Funds for the Central Universities, the National Science Foundation of China (NSFC) under Grant $61101103$ and the Scientific Research Fund of Zhejiang Provincial Education Department under Grant Y$201122655$.}

\begin{abstract}
In this paper, a recursive least squares (RLS) based blind adaptive beamforming algorithm that features a new variable forgetting factor (VFF) mechanism is presented. The beamformer is designed according to the constrained constant modulus (CCM) criterion, and the proposed adaptive algorithm operates in the generalized sidelobe canceler (GSC) structure. A detailed study of its operating properties is carried out, including a convexity analysis and a mean squared error (MSE) analysis of its steady-state behavior. The results of numerical experiments demonstrate that the proposed VFF mechanism achieves a superior learning and tracking performance compared to other VFF mechanisms.
\end{abstract}


\end{frontmatter}

\section{Introduction}
Numerous adaptive beamforming algorithms for application to wireless communication receivers have been reported in the literature in the last few decades \cite{godara1997application1,godara1997application2,RobustAdaptiveBeamforming,honig1995blind}. In this application, blind adaptation is highly desirable for the digital receivers equipped with an antenna array, since it operates without the training sequences and leads to a good solution. The constrained constant modulus (CCM) criterion \cite{gooch1986cm} is often considered as one of the most promising design criterions for blind beamforming. It takes advantage of the constant modulus (CM) property of the source modulation, while subject to a constraint on the array response to the desired user \cite{RobustAdaptiveBeamforming}, \cite{haykin2003adaptive}. The work in \cite{wang2007constrained} investigates the CCM-RLS algorithms, which combine the use of RLS adaptation with the CCM criterion, for different applications and shows that the CCM based algorithms generally outperform the ones based on constrained minimum variance (CMV).

The superior performance of the RLS-based blind beamformers is often demonstrated under the assumption of stationarity, where an ideal choice of the forgetting factor can be made. However in reality, it is difficult or even impractical to compute a predetermined value for the forgetting factor \cite{Cai2012Low}. Hence, the use of a variable forgetting factor (VFF) mechanism is an attractive choice to overcome this shortcoming. Among such mechanisms, the most common one proposed in \cite{haykin2003adaptive} is the gradient-based variable forgetting factor (GVFF), which is varied according to the measured square error at the beamformer output. Recently, the authors in \cite{Cai2012Low} have extended the conventional GVFF scheme to the CCM-RLS blind beamformer for direct sequence code division multiple access (DS-CDMA) receiver. In particular, they have proposed a new VFF mechanism that leads to a superior performance yet with a reduced complexity.

The problem formulation using the CCM criterion can be broken down into constrained and unconstrained components that give rise to the generalized sidelobe canceler (GSC) \cite{haykin2003adaptive} structure. The latter uses a main branch consisting of a fixed beamformer steered towards the desired user, in parallel with adaptive auxiliary branches that have the ability to block the signal of interest; they produce an output which ideally consists only of interference and is subtracted from that of the main branch. To the best of our knowledge, the study of effective VFF mechanisms for CCM-RLS beamformers developed around the GSC structure has not been addressed in the technical literature.

In this work, we present an extension of the method reported in \cite{Cai2012Low} for the direct-form beamformer (DFB) structure to the more practical GSC structure. The difference between these two structures has a major influence on the derivations and expressions of the adaptive weight vectors. In the GSC context, the proposed time-averaged variable forgetting factor (TAVFF) mechanism employs the time average of the CM cost function to automatically adjust the forgetting factor. Then convexity and convergence analysis of the resulting CCM-RLS algorithm with TAVFF are carried out and expressions to predict the steady-state mean squared error (MSE) are obtained. Simulation results are presented to show that the proposed TAVFF mechanism leads to a superior performance of the CCM-RLS beamformer in the GSC structure.

\section{System model and GSC beamformer design}

We consider a wireless communication scenario in which $K$ narrowband user signals impinge on a uniform linear array (ULA) comprised of $M$ identical omnidirectional antennas. Let ${{\lambda}_{c}}$ denote the wavelength and ${{d}_{s}}=\lambda_{c}/2$ be the inter-element spacing of the ULA. Assuming that the $k$th user signal impinges on the array with direction of arrival $\theta_{k}$, we can write the normalized corresponding steering vector $\mathbf{a}({{\theta}_{k}})=\frac{1}{\sqrt{M}}[1,{{e}^{-j2\pi \frac{d_s}{\lambda_c}\cos{\theta_k}}},...,{{e}^{-j2\pi \frac{d_s}{\lambda_c}\cos{\theta_k}(M-1)}}]^{T}$. Then the sampled array output vector (or snapshot) at discrete time $i \in \mathbb{N}$ can be modeled as
\begin{equation}
\mathbf{r}(i)=\mathbf{A}(\theta)\mathbf{b}(i)+\mathbf{n}(i),\quad i=0,1,2,...\label{eq:eq2}
\end{equation}
where $\mathbf{A}(\theta)=[\mathbf{a}({{\theta}_{0}}),...,\mathbf{a}({{\theta}_{K-1}})]$ is the matrix of steering vectors, $\mathbf{b}(i)=[b_0(i),\ldots,b_{K-1}(i)]^T$ is the data vector and $\mathbf{n}(i)$ is an additive vector of sensor noise with zero-mean and covariance matrix ${{\sigma }^{2}}\mathbf{I}$, where ${{\sigma}^{2}}$ denotes the variance and $\mathbf{I}$ is an identity matrix of order $M$. We assume the sequence of transmitted symbols by the desired and interference users are independent and identically distributed (i.i.d.) random processes, with values taken from a constant modulus modulation format.

The GSC structure, illustrated in Fig. \ref{fig:fig1}, converts the constrained optimization problem into an unconstrained one \cite{haykin2003adaptive}. Its output is given by $y(i)={{(v\mathbf{a}({{\theta}_{0}})-\mathbf{B}{{\mathbf{w}}}(i))}^{H}}\mathbf{r}(i),$ where $v$ is a real scalar, the signal blocking matrix $\mathbf{B}$ is orthogonal to the steering vector $\mathbf{a}({{\theta }_{0}})$ and ${\mathbf{w}}(i)$ is the complex adaptive weight vector. In this work, $\mathbf{w}(i)$ is optimized in an adaptive manner according to the CM cost function
\begin{equation}
{J_{CM}}({\mathbf{w}}(i)) = \mathbb{E}[{(|{y}(i)|^2-1)^2}].\label{eq:eq6}
\end{equation}
The CCM design have its convexity enforced by adjusting the parameter $v$, as will be discussed along with the analysis in Section \uppercase\expandafter{\romannumeral 5}. The objective of the design based on the CM cost function (\ref{eq:eq6}) is to minimize the expected deviation of the square modulus of the beamformer output from a constant while maintaining the contribution from $\theta _0$ constant, i.e., ${(v\mathbf{a}({\theta _0}) - \mathbf{B}\mathbf{w}(i))^H}\mathbf{a}({\theta _0}) = v$.

\section{Blind adaptive CCM-RLS-GSC algorithm}

For the GSC structure depicted in Fig. \ref{fig:fig1}, by employing the time-averaged estimation, we obtain the following CM cost function
\begin{equation}
\begin{aligned}
{J_{CM}}({\mathbf{w}}(i))= \sum\limits_{n=1}^i{{\lambda ^{i-n}}{{\big(|{{( {v\mathbf{a}({{\theta }_{0}})-\mathbf{B}{\mathbf{w}}(i)})}^H}\mathbf{r}(n)|^2 - 1\big)}^2}},\label{eq:eq7}
\end{aligned}
\end{equation}
where the forgetting factor $\lambda$ should be chosen as a positive constant. By taking the gradient of (\ref{eq:eq7}) with respect to $\mathbf{w}^{*}(i)$ and equating it to zero, we have
\begin{equation}
\begin{aligned}
\frac{{\partial {J_{CM}}({\mathbf{w}}(i))}}{{\partial \mathbf{w}^*}} = \sum\limits_{n=1}^{i}{{{\lambda }^{i-n}}\big(\mathbf{x}(n){{\mathbf{x}}^{H}}(n){{\mathbf{w}(i)}}-\mathbf{x}(n){{d}^{*}}(n)\big)}=\mathbf{0},\label{eq:eq9}
\end{aligned}
\end{equation}
where $\mathbf{x}(n)={{\mathbf{B}}^{H}}\tilde{\mathbf{r}}(n)$, $\tilde{\mathbf{r}}(n)=y^{*}(n)\mathbf{r}(n)$ and $d(n)=v{{\mathbf{a}}^{H}}( {{\theta }_{0}} )\tilde{\mathbf{r}}(n)-1$. Defining the correlation matrix $\mathbf{Q}(i)=\sum\limits_{n=1}^i{\lambda{(n)}^{i-n}}\mathbf{x}(n){\mathbf{x}^H}(n)$, and cross-correlation vector $\mathbf{p}(i) = \sum\limits_{n=1}^i{\lambda{(n)}^{i-n}}\mathbf{x}(n){d^*}(n)$, it follows from (\ref{eq:eq9}) that $\mathbf{w}(i) = {\mathbf{Q}^{-1}}(i)\mathbf{p}(i).$
This expression for $\mathbf{w}(i)$ has the same form as the well-known weighted least-square solution, and hence we can directly obtain the RLS equations \cite{haykin2003adaptive}
\begin{equation}
\mathbf{w}(i) = \mathbf{w}({i-1})+\mathbf{k}(i){e^*}(i),\label{eq:eq16}
\end{equation}
where $\mathbf{k}(i) = \frac{{{\mathbf{Q}^{-1}}({i-1})\mathbf{x}(i)}}{{\lambda+{\mathbf{x}^H}(i){\mathbf{Q}^{-1}}({i-1})\mathbf{x}(i)}},$ and $e(i)=d(i)-\mathbf{w}^{H}(i-1)\mathbf{x}(i).$ These equations with proper initialization define the CCM-RLS blind beamforming algorithm for the GSC structure.

\vspace{-3.5mm}

\section{Proposed TAVFF scheme}

\subsection{Blind TAVFF mechanism}
Motivated by the variable step size (VSS) mechanism for the least mean square (LMS) algorithm \cite{haykin2003adaptive} and the original work in \cite{Cai2012Low}, we introduce a new variable quantity which is updated by the instantaneous CM cost function, as follows:
\begin{equation}
\begin{aligned}
\phi(i)=\alpha \phi(i-1)+\beta{\big(|{( {v\mathbf{a}({{\theta_0}} )-\mathbf{B}{\mathbf{w}}(i)})^H}\mathbf{r}(i)|^2-1\big)^2},\label{eq:eq20}
\end{aligned}
\end{equation}
where $0<\alpha <1$ and $\beta >0$. The updated quantity $\phi(i)$ changes more or less rapidly with time and can track the average deviation from the CM property. In particular, large deviations from CM in (3) will cause $\phi(i)$ to increase, which in turn can be exploited to reduce the forgetting factor  $\lambda$ for a faster tracking, thereby alleviating the deleterious effects of sudden changes. Conversely, in the case of small deviation from CM, $\phi(i)\approx 0$ and a larger value of $\lambda$ should be employed. That is, $\lambda$ should vary in an inverse way to $\phi(i)$, while remaining within a reasonable range $0 < \lambda^- \le \lambda\le \lambda^+<1$. Using $\phi(i)$ and based on the work in \cite{Cai2012Low}, we can update $\lambda$ through the non-linear operation \footnote{As an alternative to (\ref{eq:eq21}), we have experimented with different formulas combining reciprocal, subtraction, power, etc. However, simulation results have demonstrated that the proposed formula in (\ref{eq:eq21}) can achieve the best performance among the various approaches investigated.}

\begin{equation}
\lambda(i)=\bigg[{\frac{1}{{1+\phi (i)}}}\bigg]_{\lambda^{-}}^{\lambda^{+}}\label{eq:eq21}
\end{equation}
where $[.]_{{\lambda}^{-}}^{{\lambda }^{+}}$ denotes truncation to the limits of the range $[{{\lambda }^{-}},{{\lambda}^{+}} ]$. The proposed low-complexity TAVFF is given by (\ref{eq:eq20}) and (\ref{eq:eq21}). A summary of CCM-RLS blind beamformer for the GSC structure with the TAVFF is given in Table  \uppercase\expandafter{\romannumeral 1}.

It can be shown in Table  \uppercase\expandafter{\romannumeral 2} that the computational complexity of this TAVFF mechanism has been reduced significantly compared with GVFF mechanism which is detailed in \cite{Cai2012Low}. Specifically, the GVFF requires $12M^2-12M+3$ multiplications and $5M^2-8M+5$ additions per iteration, while the proposed TAVFF only requires $5$ multiplications and $3$ additions.

\vspace{-3.5mm}

\subsection{Steady-state properties of TAVFF}
We start by considering the CM cost function where we take the additive white Gaussian noise into consideration, which has not been addressed in \cite{Cai2012Low}. Since $0<\alpha <1$, by taking the expectation of (\ref{eq:eq20}) we obtain
\begin{equation}
\mathbb{E}[{\phi(i)}] = \frac{{\beta \mathbb{E}[{{(|{{( {v\mathbf{a}( {{\theta _0}} ) - \mathbf{B}{\mathbf{w}}( i )} )}^H}\mathbf{r}( i ){|^2} - 1)}^2}]}}{{1 - \alpha }}.\label{eq:eq22}
\end{equation}
Then, we rewrite the CM cost function as follows
\begin{equation}
{J_{CM}}({\mathbf{w}}(i)) = \mathbb{E}[{(|{( {v\mathbf{a}( {{\theta _0}} ) - \mathbf{B}{\mathbf{w}}( i )} )^H}\mathbf{r}( i ){|^2} - 1)^2}] =\mathbb{E}[{(|y( i ){|^2} - 1)^2}],\label{eq:eq23}
\end{equation}
To simplify this expression, it is convenient to introduce the modified weight vector
$\tilde{\mathbf{w}}(i)=
v \mathbf{a}(\theta_0)-\mathbf{B}\mathbf{w}(i)$, in terms of which we have $y(i)=\tilde{\mathbf{w}}(i)^H \mathbf{A}(\theta)^H \mathbf{b}(i) + \tilde{\mathbf{w}}(i)^H \mathbf{n}(i)$. Since $\mathbf{b}(i)$ is an i.i.d. sequence of random vectors with zero-mean and identity covariance matrix, which is independent of the noise $\mathbf{n}(i)$, it follows according to \cite{xu2001modified} that the CM cost function can be expressed as
\begin{equation}
{{J}_{CM}}({\mathbf{w}}(i))={{J}_{1}}(\tilde{\mathbf{w}}(i))+{{\sigma }^{2}}{{J}_{2}}( {\tilde{\mathbf{w}}}(i) ),\label{eq:eq25}
\end{equation}
\begin{equation}
{{J}_{1}}(\tilde{\mathbf{w}}(i))=2{{( {{\mathbf{u}}^{H}}\mathbf{u} )}^{2}}-\sum\limits_{k=0}^{K-1}{u_{k}^{4}}-2{{\mathbf{u}}^{H}}\mathbf{u}+1,\label{eq:eq26}
\end{equation}
\begin{equation}
{{J}_{2}}( {\tilde{\mathbf{w}}}(i))=( 4{{\mathbf{u}}^{H}}\mathbf{u}-2+3{{\sigma }^{2}}{{{\tilde{\mathbf{w}}}}^{H}}\tilde{\mathbf{w}} ){{\tilde{\mathbf{w}}}^{H}}\tilde{\mathbf{w}}.\label{eq:eq27}
\end{equation}
In these expressions, vector $\mathbf{u}$ is defined as $\mathbf{u}\equiv \mathbf{u}(i)= [u_0,\ldots,.u_{K-1}]^T =\mathbf{A}(\theta)^H \tilde{\mathbf{w}}(i)$, where we have dropped the time index $i$ to simplify the notations. We assume that $\underset{i\to \infty }{\mathop{\lim }} {{J}_{1}}(\tilde{\mathbf{w}}(i))={{J}_{1}}(\tilde{\mathbf{w}}_{opt})$ is approximately equal to the steady-state noiseless value of the CM cost function where ${{\tilde{\mathbf{w}}}_{opt}} = {v\mathbf{a}({{\theta _0}})-\mathbf{B}{\mathbf{w}}_{opt}} $ denotes the optimal beamformer, and $\underset{i\to \infty }{\mathop{\lim }}{{\sigma }^{2}}{{J}_{2}}( {\tilde{\mathbf{w}}(i)})={{J}_{ex}}(\infty)$ contributes to the steady-state noisy component. Taking the high signal-to-noise-ratio (SNR) into account in our common environment \cite{Cai2012Low,xu2001modified}, we have ${{J}_{1}}(\tilde{\mathbf{w}}_{opt}) \gg {{J}_{ex}}( \infty  )$. Following the steps of analysis as in \cite{Cai2012Low}, we can obtain

\begin{equation}
\mathbb{E}[ \phi(\infty)]\approx\frac{\beta{{{J}_{1}}(\tilde{\mathbf{w}}_{opt})}}{1-\alpha}, \quad \mathbb{E}[{{\phi}^{2}}(\infty)]\approx \frac{2\alpha \beta \mathbb{E}[\phi(\infty) ]{{{J}_{1}}(\tilde{\mathbf{w}}_{opt})}}{1-{{\alpha}^{2}}}.\label{eq:eq28}
\end{equation}
Using (\ref{eq:eq21}) and with the aid of Taylor's formula, we can have the expressions of the first and second order moments of the VFF in steady-state:

\begin{equation}
\mathbb{E}[ {\lambda(\infty)}] \approx 1-\frac{{(1-{\alpha^2})\beta{{J_{1}}(\tilde{\mathbf{w}}_{opt})}-2\alpha{\beta^2}{J_{1}}(\tilde{\mathbf{w}}_{opt})^2}}{{(1-\alpha)(1- {\alpha^2})}},\label{eq:eq36}
\end{equation}

\begin{equation}
\mathbb{E}[{{\lambda }^{2}}(\infty)] \approx 1-\frac{{2(1 -{\alpha^2})\beta{{{J}_{1}}(\tilde{\mathbf{w}}_{opt})}-6\alpha{\beta^2}{{J}_{1}}(\tilde{\mathbf{w}}_{opt})^2}}{{(1-\alpha )(1-{\alpha^2})}}.\label{eq:eq37}
\end{equation}

\section{Analysis of the CCM-RLS blind GSC beamformer}

\vspace{-2mm}
\subsection{Convexity of the optimization problem}
Without loss of generality, we let user 0 be the desired user and we define $D={{u}_{0}}u_{0}^{*}={{v}^{2}}$, $\bar{\mathbf{A}}=[ \mathbf{a}( {{\theta }_{1}} ),...,\mathbf{a}( {{\theta }_{K-1}} ) ]$ and $\bar{\mathbf{u}}={{[{{u}_{1}},...,{{u}_{{K}-1}}]}^{T}}=\bar{\mathbf{A}}^{H}{\tilde{\mathbf{w}}}$. It is then possible to express the CM cost function as
\begin{equation}
{{J}_{CM}}({\mathbf{w}}(i))={{J}_{1}}({\tilde{\mathbf{w}}})+{{\sigma }^{2}}{{J}_{2}}( {\tilde{\mathbf{w}}} ),\label{eq:eq38}
\end{equation}
%where
\begin{equation}
\begin{aligned}
{{J}_{1}}({\tilde{\mathbf{w}}})= 2{{( D+{{{\bar{\mathbf{u}}}}^{H}}\bar{\mathbf{u}} )}^{2}}-({D^2} + \sum\limits_{k=1}^{K-1}{u_k^4} ) -2( D+{{{\bar{\mathbf{u}}}}^{H}}\bar{\mathbf{u}} )+1,\label{eq:eq39}
\end{aligned}
\end{equation}
\begin{equation}
{{J}_{2}}( {\tilde{\mathbf{w}}} )=( 4( D+{{{\bar{\mathbf{u}}}}^{H}}\bar{\mathbf{u}} )-2+3{{\sigma }^{2}}{{{\tilde{\mathbf{w}}}}^{H}}\tilde{\mathbf{w}} ){{\tilde{\mathbf{w}}}^{H}}\tilde{\mathbf{w}}.\label{eq:eq40}
\end{equation}
To study the convexity of ${{J}_{CM}}({\mathbf{w}}(i))$, we compute its Hessian matrix using the rule $\mathbf{H}=\frac{\partial }{\partial {{{\tilde{\mathbf{w}}}}^{H}}}\frac{\partial {{J}_{CM}}}{\partial \tilde{\mathbf{w}}}$. This yields $\mathbf{H}={{\mathbf{H}}_{1}}+{{\sigma }^{2}}{{\mathbf{H}}_{2}}$, where
\begin{equation}
{\mathbf{H}_1} = 4\bar {\mathbf{A}}[(D - 1/2)\mathbf{I} + {{\bar {\mathbf{u}}}^H}\bar {\mathbf{u}}\mathbf{I} + \bar {\mathbf{u}}{{\bar {\mathbf{u}}}^H} - \text{diag}({| {{u_1}} |^2},...,{| {{u_{K-1}}} |^2})]{{\bar {\mathbf{A}}}^T},\label{eq:eq41}
\end{equation}
\begin{equation}
\begin{aligned}
&{{\mathbf{H}}_{2}}=( 4D-2 )\mathbf{I}+6{{\sigma }^{2}}( {{{\tilde{\mathbf{w}}}}^{H}}\tilde{\mathbf{w}}\mathbf{I}+\tilde{\mathbf{w}}{{{\tilde{\mathbf{w}}}}^{H}} ) + 4({{\tilde {\mathbf{w}}}^H}\bar {\mathbf{A}}{{\bar {\mathbf{A}}}^H}\tilde {\mathbf{w}}\mathbf{I} + {(\bar {\mathbf{A}}{{\bar {\mathbf{A}}}^H})^T}{{\tilde {\mathbf{w}}}^H}\tilde {\mathbf{w}}\\ & + {(\tilde {\mathbf{w}}{{\tilde {\mathbf{w}}}^H}\bar {\mathbf{A}}{{\bar {\mathbf{A}}}^H})^T} + {({{\tilde {\mathbf{w}}}^H}\bar {\mathbf{A}}{{\bar {\mathbf{A}}}^H}\tilde {\mathbf{w}})^T}).\label{eq:eq42}
\end{aligned}
\end{equation}
For ${{\mathbf{H}}_{1}}$ in (\ref{eq:eq41}), the second, third and fourth terms yield the positive definite matrix $4\bar {\mathbf{A}}\big( \bar{\mathbf{u}}{{{\bar{\mathbf{u}}}}^{H}}+\text{diag}(\sum\limits_{k=2}^{K-1} {|{{u}_{k}}|^{2}},...,\sum\limits_{k=1,k{\neq}K-1}^{K-1} {|{{u}_{k}}|^{2}} )\big){{\bar {\mathbf{A}}}^T}$, while the first term provides the condition $D={{v}^{2}}\ge 1/2$, which ensures the convexity of ${{J}_{1}}({\tilde{\mathbf{w}}})$. Since $\mathbf{H}_2$ in (\ref{eq:eq42}) is a smooth, differentiable function of $\tilde{\mathbf{w}}$, it follows that for small values of $\sigma^2$, the cost function $J_{CM}(\mathbf{w}(i))$ in (\ref{eq:eq38}) remains convex under  perturbation from the noise-free case by the term $\sigma^2$ $J_2(\tilde{\mathbf{w}})$ \cite{xu2001modified}. For larger values of $\sigma^2$, the constant $v$ can be adjusted to a sufficiently large value such that $\mathbf{H}_2$ is positive definite in any bounded region. We conclude that by properly selecting $v$, $\mathbf{H}$ can be made positive definite, this implies that the cost function $J_{CM}(\mathbf{w}(i))$ is strictly convex and therefore, the algorithm can reach the global minimum.

\vspace{-2mm}
\subsection{Convergence of the mean weight vector}
The weight vector update equation ${{\mathbf{w}}}( i )={{\mathbf{w}}}( i-1 )+\mathbf{k}( i ){{e}^{*}}( i )$, has the same form as in the well-known RLS solution, which makes it convenient to analyze the convergence performance. This similarity results from employing the GSC structure instead of DFB structure as in \cite{Cai2012Low}. Before taking the analysis further, we should note that while the input data in the conventional RLS algorithm is the array output vector $\mathbf{r}(i)$, in our proposed CCM-RLS algorithm the input data is $\mathbf{x}(i)=\mathbf{B}^{H}y^{*}(i)\mathbf{r}(i)$. Despite this difference, we can still employ the principle of orthogonality between the optimum error and the data, which give $\mathbb{E}[ \mathbf{x}(i)e_{opt}^{*}(i)]=\mathbf{0}$, where ${e_{opt}}(i)=d(i)-\mathbf{w}_{opt}^{H}\mathbf{x}(i)$ denotes the optimum error. By following the convergence analysis of the mean weight vector for the RLS solution and recalling that $0<\mathbb{E}[\lambda(i)]<1$, we can finally obtain
\begin{equation}
\underset{i\to \infty }{\mathop{\lim }}\mathbb{E}[{{\mathbf{w}}}( i )-{{\mathbf{w}}_{opt}}]=0.\label{eq:eq53}
\end{equation}
This shows that the expected weight error converges to zero as $i\to \infty$.

\subsection{Convergence of MSE}
Next we discuss the convergence of the MSE for the proposed CCM-RLS algorithm and provide an analytical expression to predict its steady-state value. It can be shown that the steady-state MSE is given by
\begin{equation}
\begin{aligned}
\underset{i\to \infty }{\mathop{\lim}}{{\xi}_{mse}}(i)&=\underset{i\to \infty}{\mathop{\lim}}\mathbb{E}[{{|{{b}_{0}}(i)-{{{\tilde{\mathbf{w}}}}^{H}}(i)\mathbf{r}(i)|}^{2}}] \\
&=(1-2v)+\mathbb{E}[\tilde{\mathbf{w}}_{opt}^{H}\mathbf{R}(i){{{\tilde{\mathbf{w}}}}_{opt}}] +\underset{i\to \infty}{\mathop{\lim}}\mathbb{E}[tr[\mathbf{R}(i)\tilde{\mbox{\boldmath$\varepsilon$}}(i){{{\tilde{\mbox{\boldmath$\varepsilon$}}}}^{H}}(i)]], \label{eq:eq54}
\end{aligned}
\end{equation}
where $\tilde{\mbox{\boldmath$\varepsilon$}}(i)=\tilde{\mathbf{w}}(i)-{{\tilde{\mathbf{w}}}_{opt}}$. We shall define $\mathbf{\Theta}(i)=\mathbb{E}[ \tilde{\mbox{\boldmath$\varepsilon$}}(i){{{\tilde{\mbox{\boldmath$\varepsilon$}}}}^{H}}(i)]$ and neglect the dependence among $e_{opt}(i)$, ${{\mathbf{x}}}(i)$ and ${{\mathbf{Q}}^{-1}}(i)$ in the limit as $i\to \infty $ \cite{haykin2003adaptive}. We obtain
\begin{equation}
\mathbf{\Theta}(i)=\mathbb{E}[{{\lambda}^{2}}(i)]\mathbf{\Theta}(i-1) +\sigma_{opt}^{2}\mathbf{B}\mathbb{E}[{{\mathbf{Q}}^{-1}}(i)]\mathbb{E}[\mathbf{x}(i){{\mathbf{x}}^{H}}(i)]\mathbb{E}[{{\mathbf{Q}}^{-1}}(i)]{{\mathbf{B}}^{H}},  \label{eq:eq57}
\end{equation}
where $\sigma_{opt}^2=\mathbb{E}[|e_{opt}(i)|^2]$. It converges since $\lambda(i)<1$. At steady-state we have
\begin{equation}
\underset{i\to \infty }{\mathop{\lim }}\mathbf{\Theta}(i)\approx\frac{\sigma_{opt}^{2}(1-\mathbb{E}[ \lambda{(\infty)}])^{2}}{(1-\mathbb{E}[ {{\lambda}^{2}}(\infty)])\mathbb{E}[{{|{y}_{opt}(i) |}^{2}} ]}\mathbf{B}{{( {{\mathbf{B}}^{H}}\mathbf{R}\mathbf{B} )}^{-1}}{{\mathbf{B}}^{H}}, \label{eq:eq63}
\end{equation}
where ${y_{opt}}(i)=\tilde{\mathbf{w}}_{opt}^{H}\mathbf{r}(i)$, ${\mathbf{R}}=\mathbb{E}[\mathbf{r}(i){{\mathbf{r}}^{H}}(i)]$, and $\mathbb{E}[{{\lambda}}(\infty)]$ and $\mathbb{E}[{{\lambda}^{2}}(\infty) ]$ are given in (\ref{eq:eq36}) and (\ref{eq:eq37}).
Finally, based on (\ref{eq:eq54}) and (\ref{eq:eq63}), we can derive the expressions to predict the steady-state MSE.

\vspace{-4.5mm}

\section{Simulations}
\vspace{-2mm}
In this section, we evaluate the performance of the proposed TAVFF with the blind adaptive CCM-RLS-GSC beamformer. On the TX side, the BPSK modulation is employed to keep the CM property, and the source power is normalized. On the RX side, the uniform linear array is composed of $M=16$ sensor elements. In our simulations, an experiment is made up of 1000 independent runs and for each such experiment, the DOAs of the users are randomly generated with a uniform distribution between 0 and 180 degrees and kept fixed for all the runs. The exact DOA of the desired user is assumed to be known by the beamformer. There are $K=5$ users including the desired one and four interferers in a static environment and the input SNR is fixed at 15dB unless otherwise indicated. The performance is not sensitive to the initial values,
accordingly, we set $v=1$, $\mathbf{Q}^{-1}(0)=\delta^{-1}\mathbf{I}\footnote{$\delta=10$ for $SNR=0$dB, $\delta=0.1$ for $SNR=20$dB and $\delta=1$ for $SNR=5,10,15$dB.}$,  generate $\mathbf{w}(0)$ randomly.

The result in Fig. \ref{fig:fig2} (a) show that as the number of snapshots increases, the MSE value converges and reaches a steady-state level which is in good agreement with the analytical result given by (\ref{eq:eq54}). In Fig. \ref{fig:fig2} (b), we further compare the simulation and theoretical results by showing the steady-state MSE versus SNR. We compute the MSE value at snapshot $1000$ for each run, and $1000$ independent runs are averaged to get the final result for each exact SNR. We find that the results agree well with each other over the considered SNR range.

In the experiment of Fig. \ref{fig:fig3}, we evaluate the beamformer SINR performance against the number of snapshots. In all cases, the output SINR values increase to the steady-state as the number of snapshots increases. The graph illustrates that the CCM-RLS-GSC beamformer with the proposed TAVFF achieves the fastest convergence and the best performance. We list the observed SINR value at snapshot $800$ in the steady-state regime in Table \uppercase\expandafter{\romannumeral 3}, where the results show that the performance improvements with TAVFF are statistically significant.

In Fig. 4, we assess the SINR performance of the proposed TAVFF and fixed forgetting factor against the number of snapshots for both the DFB and GSC structures. The resulting curves show that for both structures, the performance of the CCM-RLS algorithm with TAVFF is significantly better than the corresponding with fixed forgetting factor. The GSC structure leads to an improved performance compared to the DFB, especially a much faster initial convergence. The observed steady-state SINR values at snapshot $800$ are listed in Table  \uppercase\expandafter{\romannumeral 4}.

At last, we evaluate the SINR convergence performance in a nonstationary scenario. The system starts with $K=5$ users, after $1000$ snapshots, two more interferers having the same power enter the system. From the results in Fig. \ref{fig:fig5}, we can see the abrupt change at $1000$ snapshot reduces the output SINR and degrades the performance of all the algorithms. However, TAVFF can quickly track this change and recover faster to a new and larger steady-state. To make this clear, we have measured the rates of convergence (in dB per iteration) and SINR standard deviation of the various algorithms over time. The results, listed in Table \uppercase\expandafter{\romannumeral 5} and \uppercase\expandafter{\romannumeral 6}, show that while all the presented algorithms recover from the newly introduced interference, TAVFF mechanism exhibits the fastest convergence rate, smallest standard deviation and the best performance.
%in such a non-stationary environment
%}}

\vspace{-4.5mm}

\section{Conclusion}
In this paper, we developed a CCM-RLS-based blind adaptive beamforming algorithm that features a new proposed low-complexity TAVFF mechanism and operates within the GSC structure for its realization. The convergence properties of this algorithm were analyzed, including the study of convexity and steady-state MSE behavior. The simulation results were in good agreement with the theoretical ones and showed that the proposed TAVFF mechanism outperforms other VFF mechanisms previously developed for CCM-RLS blind adaptive beamforming.

%%%%%%%%%%%%%%%%%%%%%%%%%%%%%%%%%%%%%%%%%%%%%%

\newpage
\renewcommand{\thetable}{\Roman{table}}
\renewcommand{\thefigure}{\arabic{figure}}

%%%%%%%%%%%%%table 1%%%%%%%%%%%%%%%%%%%%%%%%%
\begin{table}[p]
\begin{center}
\caption{The CCM-RLS-GSC-TAVFF algorithm}
\begin{tabular}{l}
\hline
Initialization:\\
\vspace{1mm}
Initialize ${{\mathbf{Q}}^{-1}}(0)$, ${{\mathbf{w}}}(0)$.\\

\hline
Update for each time index $i$\\
\vspace{1mm}
Coefficient updating:\\
\vspace{1mm}
$\tilde{\mathbf{w}}( i-1 )=v\,\mathbf{a}( {{\theta }_{0}} )-\mathbf{B}{{\mathbf{w}}}( i-1 )$, $y( i )={{\tilde{\mathbf{w}}}^{H}}( i-1 )\mathbf{r}( i ),$\\
\vspace{1mm}
$\tilde{\mathbf{r}}( i )=y^{*}( i )\mathbf{r}( i )$, $\mathbf{x}( i )={{\mathbf{B}}^{H}}\tilde{\mathbf{r}}( i )$, $d( i )=v\,{{\mathbf{a}}^{H}}( {{\theta }_{0}} )\tilde{\mathbf{r}}( i )-1,$\\
\vspace{1mm}

Adaptation gain computation:\\
$\mathbf{k}( i ) = \frac{{{\mathbf{Q}^{ - 1}}( {i - 1} )\mathbf{x}( i )}}{{\lambda  + {\mathbf{x}^H}( i ){\mathbf{Q}^{ - 1}}( {i - 1} )\mathbf{x}( i )}},$\\
\vspace{1mm}
${{\mathbf{Q}}^{-1}}( i )={{\lambda }^{-1}}( i-1 ){{\mathbf{Q}}^{-1}}( i-1 )-{{\lambda }^{-1}}( i-1 )\mathbf{k}( i ){{\mathbf{x}}^{H}}( i ){{\mathbf{Q}}^{-1}}( i-1 ),$\\
\vspace{1mm}

Forgetting factor updating:\\
\vspace{1mm}
$\phi ( i )=\alpha \phi ( i-1 )+\beta {{( {{| y( i ) |}^{2}}-1 )}^{2}},$\\
\vspace{1mm}
$\lambda ( i ) = \big[ {\frac{1}{{1 + \phi ( i )}}} \big]_{{\lambda ^ - }}^{{\lambda ^ + }},$\\
\vspace{1mm}

Weight vector calculation:\\
\vspace{1mm}
$e( i )=d( i )-{\mathbf{w}}( i-1 )\mathbf{x}( i ),$\\
\vspace{1mm}
${{\mathbf{w}}}( i )={{\mathbf{w}}}( i-1 )+\mathbf{k}( i ){{e}^{*}}( i ).$\\

\hline
\end{tabular}
\label{tab:tab1}
\end{center}
\end{table}

\begin{table}[h]
\begin{center}
\caption{Additional Computational Complexity}
\begin{tabular}{l}
\hline
\quad\quad\quad\quad\quad\quad\quad\quad Number of operations per symbol\\

\hline
\textbf{Mechanism} \quad\quad\quad\quad\quad\quad multiplications \quad\quad\quad\quad\quad\quad   additions\\

\hline
\textbf{TAVFF} \quad\quad\quad\quad\quad\quad\quad\quad\quad\quad\quad$5$ \quad\quad\quad\quad\quad\quad\quad\quad\quad\quad  $3$\\

\hline
\textbf{blind GVFF} \quad\quad\quad\quad\quad$12{M^2} - 12M + 3$ \quad\quad\quad\quad $5{M^2} - 8M + 5$\\

\hline
\end{tabular}
\label{tab2}
\end{center}
\end{table}

\begin{table}[h]
\begin{center}
\caption{SINR steady-state values in Fig. \ref{fig:fig3}}
\begin{tabular}{l}

\hline
\textbf{Algorithm} \quad\quad\quad\quad SINR(dB, 95\% confidence interval) \\

\hline
\textbf{CMV} \quad\quad\quad\quad\quad\quad\quad\quad\quad\quad$10.31\pm0.05$ \\

\hline
\textbf{CCM} \quad\quad\quad\quad\quad\quad\quad\quad\quad\quad$10.84\pm0.04$ \\

\hline
\textbf{CCM-GVFF} \textbf{{\color{white}{0}}}\quad\quad\quad\quad\quad\quad$11.68\pm0.02$ \\

\hline
\textbf{CCM-TAVFF} \quad\quad\quad\quad\quad\quad$12.19\pm0.01$ \\

\hline
\end{tabular}
\label{tab3}
\end{center}
\end{table}

\begin{table}[h]
\begin{center}
\caption{SINR steady-state values in Fig. \ref{fig:fig4}}
\begin{tabular}{l}

\hline
\textbf{Algorithm} \quad\quad\quad\quad SINR (dB, 95\% confidence interval) \\

\hline
\textbf{DFB} \quad\quad\quad\quad\quad\quad\quad\quad\quad\quad$10.22\pm0.05$ \\

\hline
\textbf{GSC} \quad\quad\quad\quad\quad\quad\quad\quad\quad\quad$10.80\pm0.03$ \\

\hline
\textbf{DFB-TAVFF} \quad\quad\quad\quad\quad\quad$11.70\pm0.03$ \\

\hline
\textbf{GSC-TAVFF} \quad\quad\quad\quad\quad\quad$12.21\pm0.01$ \\

\hline
\end{tabular}
\label{tab4}
\end{center}
\end{table}

\begin{table}[h]
\begin{center}
\caption{SINR Convergence Rate (dB/iteration)}
\begin{tabular}{l}

\hline
\textbf{snapshot}  \quad\quad\quad {CMV} \quad\quad\quad\quad {CCM} \quad\quad {CCM-GVFF} \quad\quad {CCM-TAVFF}\\

\hline
\quad\textbf{{\color{white}{000}}{0}} \quad\quad\quad\quad ${5.341}$ \quad\quad\quad\quad ${6.512}$ \quad\quad\quad\quad ${7.510}$ \quad\quad\quad\quad\quad ${8.488}$ \\

\hline
\quad\textbf{{\color{white}{00}}{40}} \quad\quad\quad\quad ${0.158}$ \quad\quad\quad\quad ${0.265}$ \quad\quad\quad\quad ${0.272}$ \quad\quad\quad\quad\quad ${0.278}$ \\

\hline
\quad\textbf{{\color{white}{00}}{80}} \quad\quad\quad\quad ${0.094}$ \quad\quad\quad\quad ${0.134}$ \quad\quad\quad\quad ${0.138}$ \quad\quad\quad\quad\quad ${0.143}$ \\

\hline
\quad\textbf{{\color{white}{0}}{120}} \quad\quad\quad\quad ${0.069}$ \quad\quad\quad\quad ${0.091}$ \quad\quad\quad\quad ${0.093}$ \quad\quad\quad\quad\quad ${0.096}$ \\

\hline
\quad\textbf{1000} \quad\quad\quad\quad ${6.346}$ \quad\quad\quad\quad ${8.094}$ \quad\quad\quad\quad ${8.545}$ \quad\quad\quad\quad\quad ${9.868}$ \\

\hline
\quad\textbf{1040} \quad\quad\quad\quad ${0.221}$ \quad\quad\quad\quad ${0.237}$ \quad\quad\quad\quad ${0.248}$ \quad\quad\quad\quad\quad ${0.268}$ \\

\hline
\quad\textbf{1080} \quad\quad\quad\quad ${0.118}$ \quad\quad\quad\quad ${0.124}$ \quad\quad\quad\quad ${0.129}$ \quad\quad\quad\quad\quad ${0.138}$ \\

\hline
\quad\textbf{1120} \quad\quad\quad\quad ${0.081}$ \quad\quad\quad\quad ${0.085}$ \quad\quad\quad\quad ${0.088}$ \quad\quad\quad\quad\quad ${0.093}$ \\

\hline
\end{tabular}
\label{tab5}
\end{center}
\end{table}

\begin{table}[h]
\begin{center}
\caption{SINR Standard Deviation (dB)}
\begin{tabular}{l}

\hline
\textbf{snapshot}  \quad\quad\quad CMV \quad\quad\quad\quad CCM \quad\quad CCM-GVFF \quad\quad CCM-TAVFF\\

\hline
\quad\textbf{{\color{white}{0}}400} \quad\quad\quad\quad $0.89$ \quad\quad\quad\quad\quad $0.49$ \quad\quad\quad\quad\quad $0.40$ \quad\quad\quad\quad\quad $0.21$ \\

\hline
\quad\textbf{{\color{white}{0}}600} \quad\quad\quad\quad $0.88$ \quad\quad\quad\quad\quad $0.49$ \quad\quad\quad\quad\quad $0.40$ \quad\quad\quad\quad\quad $0.17$ \\

\hline
\quad\textbf{{\color{white}{0}}800} \quad\quad\quad\quad $0.87$ \quad\quad\quad\quad\quad $0.49$ \quad\quad\quad\quad\quad $0.40$ \quad\quad\quad\quad\quad $0.14$ \\

\hline
\quad\textbf{{\color{white}{0}}999} \quad\quad\quad\quad $0.86$ \quad\quad\quad\quad\quad $0.48$ \quad\quad\quad\quad\quad $0.39$ \quad\quad\quad\quad\quad $0.13$ \\

\hline
\quad\textbf{1400} \quad\quad\quad\quad $0.86$ \quad\quad\quad\quad\quad $0.53$ \quad\quad\quad\quad\quad $0.44$ \quad\quad\quad\quad\quad $0.20$ \\

\hline
\quad\textbf{1600} \quad\quad\quad\quad $0.82$ \quad\quad\quad\quad\quad $0.53$ \quad\quad\quad\quad\quad $0.44$ \quad\quad\quad\quad\quad $0.18$ \\

\hline
\quad\textbf{1800} \quad\quad\quad\quad $0.80$ \quad\quad\quad\quad\quad $0.53$ \quad\quad\quad\quad\quad $0.44$ \quad\quad\quad\quad\quad $0.17$ \\

\hline
\quad\textbf{1999} \quad\quad\quad\quad $0.79$ \quad\quad\quad\quad\quad $0.52$ \quad\quad\quad\quad\quad $0.43$ \quad\quad\quad\quad\quad $0.17$ \\

\hline
\end{tabular}
\label{tab6}
\end{center}
\end{table}

\newpage
%%%%%%%%%%%%%figure 1%%%%%%%%%%%%%%%%%%%%%%%%%
\begin{figure}[p]
\centering
\scalebox{1.5}{\includegraphics{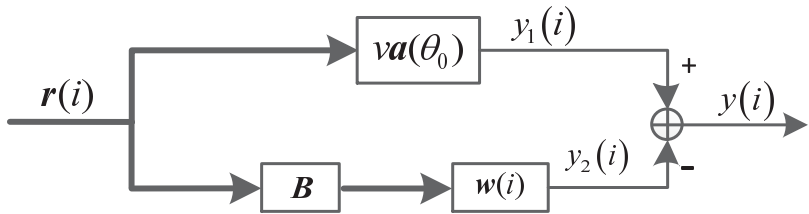}}
\caption{GSC structure for blind beamforming} \label{fig:fig1}
\end{figure}

%%%%%%%%%%%%%figure 2%%%%%%%%%%%%%%%%%%%%%%%%%
\begin{figure}[p]
  \centering
  \includegraphics[width = 0.7\textwidth]{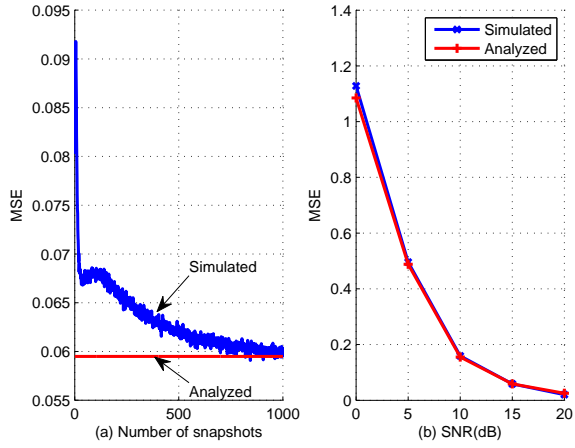}\\
  \caption{Analytical MSE versus simulated performance for the CCM-RLS-GSC algorithm with the TAVFF mechanism (number of users $K=5$, input $SNR = 15$dB).}
  \label{fig:fig2}
\end{figure}

%%%%%%%%%%%%%figure 3%%%%%%%%%%%%%%%%%%%%%%%%%
\begin{figure}[p]
  \centering
  \includegraphics[width = 0.7\textwidth]{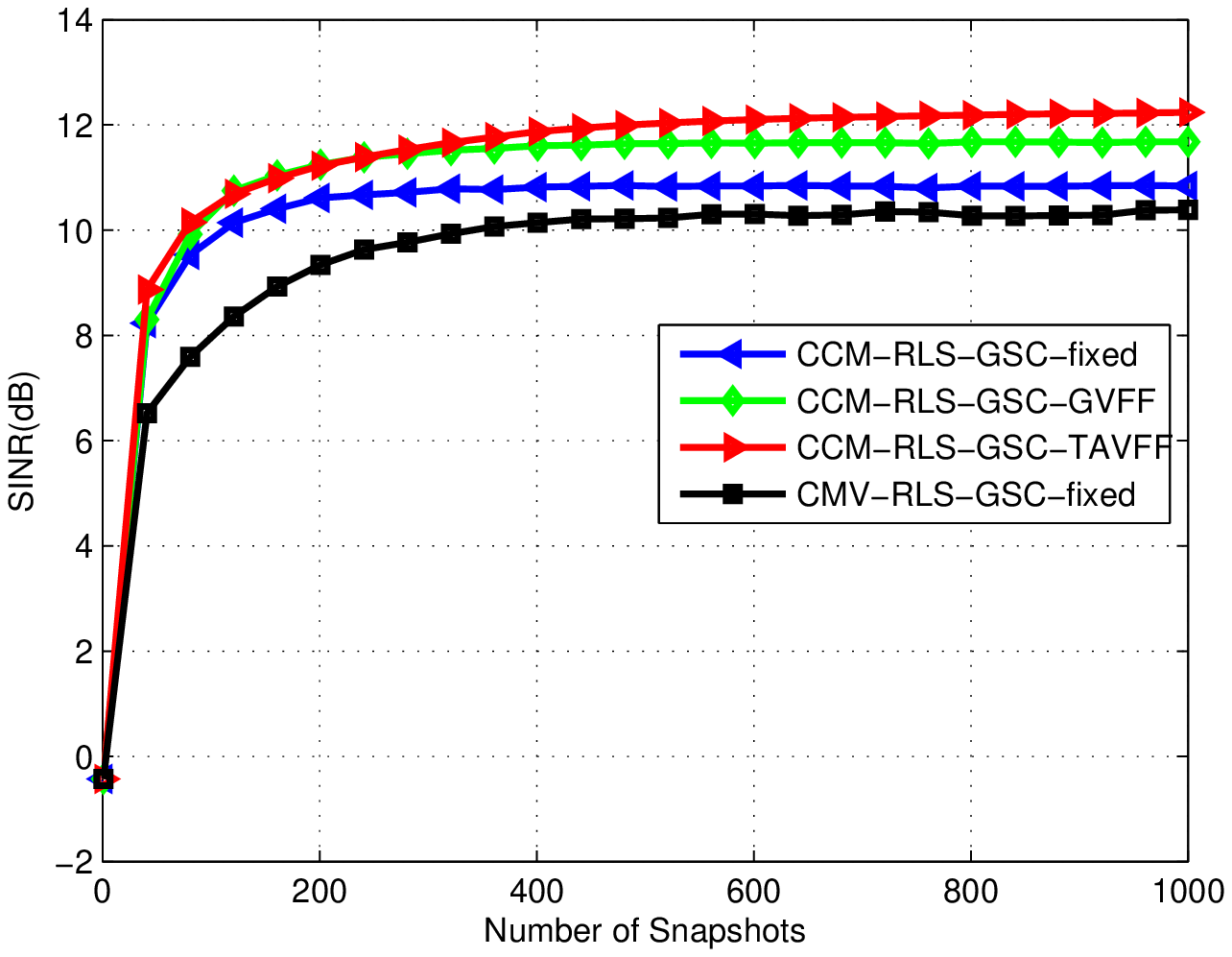}\\
  \caption{Output SINR against the number of snapshots (number of users $K=5$, input $SNR = 15$dB).}
  \label{fig:fig3}
\end{figure}

%%%%%%%%%%%%%figure 4%%%%%%%%%%%%%%%%%%%%%%%%%
\begin{figure}[p]
  \centering
  \includegraphics[width = 0.7\textwidth]{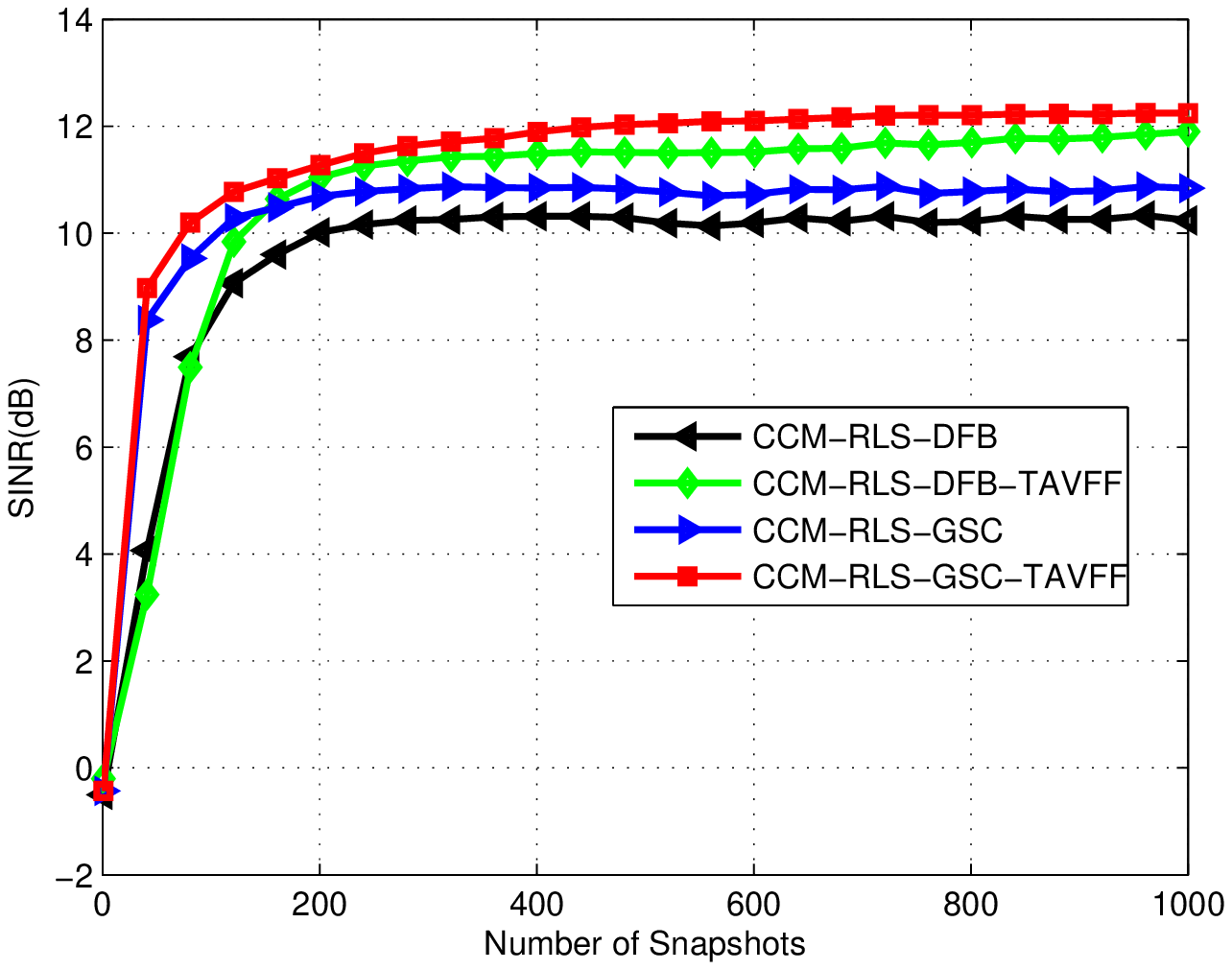}\\
  \caption{Output SINR against the number of snapshots (number of users $K=5$, input $SNR = 15$dB).}
  \label{fig:fig4}
\end{figure}

%%%%%%%%%%%%%figure 5%%%%%%%%%%%%%%%%%%%%%%%%%
\begin{figure}[p]
  \centering
  \includegraphics[width = 0.7\textwidth]{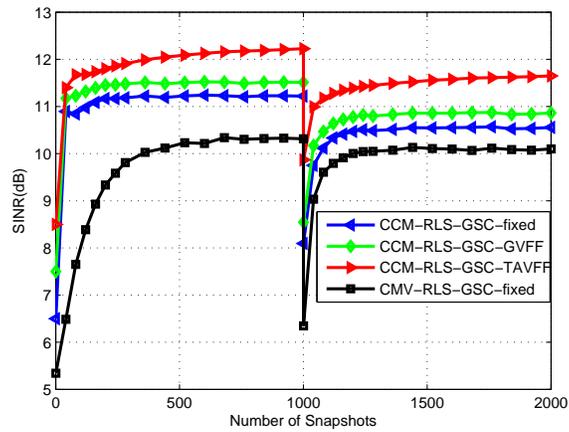}\\
  \caption{SINR performance in a nonstationary environment (For the first stage, number of users is $K=5$, for the second stage, number of users is $K=7$, input $SNR = 15$dB).}
  \label{fig:fig5}
\end{figure}

\end{document}